\documentclass[fleqn,usenatbib]{mnras}

\usepackage{newtxtext,newtxmath}

\usepackage[T1]{fontenc}

\DeclareRobustCommand{\VAN}[3]{#2}
\let\VANthebibliography\thebibliography
\def\thebibliography{\DeclareRobustCommand{\VAN}[3]{##3}\VANthebibliography}

\usepackage{graphicx}	
\usepackage{amsmath}	
\usepackage{xcolor}

\title[GC + UCD Distances]{Distance Estimation using Globular Clusters and Ultra-Compact Dwarfs}

\author[B. van Heumen et al.]{
Bas van Heumen,$^{1}$\thanks{E-mail: bvanheumen@swin.edu.au}
Jonah S. Gannon,$^{2,3,1}$
Duncan A. Forbes,$^{1}$
Jean P. Brodie$^{1}$ and
Aaron J. Romanowsky,$^{4,5}$
\\
$^{1}$Centre for Astrophysics and Supercomputing, Swinburne University, John Street, Hawthorn VIC 3122, Australia\\
$^{2}$Department of Astronomy and Astrophysics, University of Toronto, 50 St. George Street, Toronto, ON M5S 3H4, Canada\\
$^{3}$Dragonfly Focused Research Organization, 150 Washington Avenue, Suite 201, Santa Fe, NM 87501, USA\\
$^{4}$Department of Physics \& Astronomy, San Jos\'e State University, One Washington Square, San Jose, CA 95192, USA\\
$^{5}$Department of Astronomy \& Astrophysics, University of California Santa Cruz, 1156 High Street, Santa Cruz, CA 95064, USA
}

\date{Accepted XXX. Received YYY; in original form ZZZ}

\pubyear{\the\year{}}

\begin{document}
\label{firstpage}
\pagerange{\pageref{firstpage}--\pageref{lastpage}}
\maketitle

\begin{abstract}
The tight scaling relation between absolute magnitude and internal velocity dispersion ($\sigma$) for globular clusters (GCs), referred to as the GC velocity dispersion (GCVD) relation, has shown great potential as a distance estimator. However, at the high-luminosity end of the GC luminosity function, GCs mix with ultra-compact dwarfs (UCDs). As GCs appear to smoothly transition into UCDs with dynamical mass, we investigate the possibility of a unified distance estimation relation applicable to both GCs and UCDs. To this end, we exploit the transition between scaling relations with dynamical mass, using the Milky Way and M31 GCs alongside literature UCDs. Additionally, we look at the influence of UCDs, and of GC size, on GCVD-derived distances. Using the GCVD for UCDs gives a systematic distance underestimation of $\sim33$ per cent, but a cut to select UCDs with GC-like sizes can eliminate this bias. The minor systematics on GC size produce systematic offsets smaller than the uncertainty of the GCVD. Using the bilinear relation formed by the GCs and UCDs in the absolute magnitude -- log dynamical mass plane for distance estimation yields per-object distance uncertainties of $\sim$30–35 per cent (0.12–0.16 dex). We find that the GC -- UCD division occurs at $M_V \simeq -10.8$ mag and $M_{\rm dyn} \simeq 3.1\times10^6$ $M_{\odot}$, consistent with previous findings. Applied to the dwarf galaxy NGC1052-DF2, whose distance is strongly debated, the relation returns 19.1 $\pm$ 4.3 Mpc, consistent with literature values. 
\end{abstract}




\section{Introduction} \label{sec:introduction}
Globular clusters (GCs) show a strong correlation between their luminosity and internal velocity dispersions \citep[$\sigma$; e.g.][]{Pryor1993,Djorgovski1994}, as a consequence of the Virial theorem. In the same way that the Faber -- Jackson relation can be used for distance estimation \citep[e.g.][]{Faber1976} to elliptical galaxies by estimating the absolute magnitude from $\sigma$, so can the $M_V-\sigma$ relation for GCs be used to estimate distances to GCs and in turn to their host galaxy. As GCs broadly display similar properties across galaxy types and are typically numerous \citep[e.g.][]{Brodie2006}, the GC velocity dispersion (GCVD) relation has the potential for great accuracy as a galaxy distance estimator since the ensemble of GCs drives down the statistical uncertainty. \cite{Paturel1992} were the first to explore this application of the $M_V-\sigma$ correlation for GCs and found it to be consistent within 0.2 mag for the distance moduli (or $\sim9$\% for the distance) of M31 and the LMC, with only 18 Galactic GCs for calibration. This application of the GCVD has only recently been revisited by \cite{Beasley2024}, who showed that even with single-digit numbers of GCs, the distance derived from the GCVD method was consistent with the median of literature estimates within 10\%.

With a characteristic GC luminosity of $M_V \sim -7.5$ \citep[e.g.][]{Beasley2020}, current instrumentation, such as fibre-fed multiplex spectrographs, allows the GCVD to be readily applied to nearby galaxies and even out $\sim 20$ Mpc with tens of hours of observing time investment \citep[e.g.][]{Beasley2025,Fahrion2026}. As the GCVD is accurate even with only a limited sample of GCs, it suggests that the GCVD will perform well even at further distances when limited to only observing the brightest objects. However, in this regime GCs and ultra-compact dwarfs (UCDs; \citealt{Drinkwater2000,Phillipps2001}) become difficult to distinguish. With 2D half-light radii ($R_h$) of $7$--$100$~pc, luminosities of $-10 \geq M_V \geq -13$~mag, and elevated mass-to-light ratios with respect to GCs, UCDs fill the parameter space between GCs and compact galaxies \citep[e.g.][]{Dabringhausen2008,Misgeld2011,Norris2014}. As such, it is thought that UCDs are a composite population of the highest mass or merged GCs \citep[e.g.][]{Fellhauer2002,Forbes2011,Mieske2012} and stripped nuclei of now destroyed dwarf galaxies \citep[e.g.][]{Bekki2003,Pfeffer2013,Wang2023}. The relative prominence of each formation channel, however, remains under debate \citep[e.g.][]{DaRocha2011,Brodie2011,Norris2014,Pfeffer2014}. 

UCDs are commonly found to lie off the GCVD relation \citep[e.g.][]{Mieske2008}, being systematically brighter at fixed $\sigma$ due to their larger sizes and correspondingly higher required dynamical masses. This results in an underestimation of GCVD-derived distances for UCDs. Owing to their comparatively larger size, UCDs can be effectively excluded when included in a GC sample. However, this is not always possible with a limited sample.

It has been established that UCDs follow an $M/L$ -- mass relation which smoothly emerges from the range of $M/L$ values measured for GCs above a specific mass threshold: $M_{\rm dyn} \geq 2\times10^6$~$M_{\odot}$ \citep{Hasegan2005,Mieske2008}. This mass-dependent break points to the possibility of using, instead of excluding, UCDs for distance estimations through a unified relation in the luminosity -- dynamical mass plane. While masses for UCDs are typically obtained from mass modelling \citep[e.g.][]{Hilker2007}, the majority of the UCDs, like GCs, are well fitted by a general King profile, meaning the quantity $\sigma^2R_h$ can serve as a proxy for dynamical mass. 

Beyond the benefit of being applicable to larger distances, since UCDs extend up to 3 magnitudes in luminosity beyond the brightest GCs, dynamical mass should be a more direct tracer of stellar mass than $\sigma$ alone and should therefore be capable of producing more accurate estimates. Additionally, the inclusion of $R_h$ through dynamical mass will also prevent any biases through systematic differences in $R_h$ that may be present in samples of objects \citep[e.g.][]{Larsen2001,Jordan2005,Georgiev2009}.

In this paper, we investigate the impact of UCDs and $R_h$ systematics on GCVD-derived distances. Additionally, we look at the accuracy of a unified relation for GCs and UCDs for distance estimation using data from the Milky Way (MW) and M31 GCs alongside literature UCDs. We apply that relation to the unique case of the dwarf galaxy NGC1052-DF2.

\section{Data} \label{sec:data}
We use the data from the `Structural Parameters of Galactic Globular Clusters' table available online\footnote{\url{https://people.smp.uq.edu.au/HolgerBaumgardt/globular/parameter.html}} for the MW GCs. This table assembles the results of a series of studies on the Galactic GCs using fits to a suite of N-body simulations \citep[e.g.][]{Baumgardt2018}. In particular, we use the GC distance, total dynamical mass, apparent $V$-band magnitude and 3D half-mass and $R_h$. In addition we include the $E(B-V)$ values from the \citet[][2010 edition]{Harris1996} catalogue. We exclude any GC with $E(B-V) \geq 3$ to avoid the extra uncertainty associated with heavily obscured GCs. The apparent $V$-band magnitudes are extinction-corrected and converted to absolute $V$-band magnitudes using the GC distances from the table from \cite{Baumgardt2021}. Following \cite{Beasley2024}, we use the Virial theorem to calculate the global velocity dispersion, $\sigma_\infty$, using the GC total mass as its virial mass and the 3D half-mass radius. We similarly remove any GCs with $\sigma_{\infty} < 1.5$ km s$^{-1}$ to be consistent with \cite{Beasley2024}, resulting in a sample of 112 MW GCs.

For M31 GCs, we use the study of \cite{Strader2011}. They reported global velocity dispersions, extinction-corrected $V$-band luminosities and King profile structural parameters for a sample of 163 M31 GCs under the assumption of a distance to M31 of 780 kpc. While M31 GCs have been found a projected distance of 150 kpc away, the majority of the M31 GC system lies with 20 kpc \citep[e.g.][]{Huxor2014}. We use their reported $\sigma_\infty$ and $V$-band luminosities from their tables 4, 5 and 6. This sample has a minimum global velocity dispersion of $\sigma_\infty = 2.1$ km s$^{-1}$ and none of the GCs have $E(B-V) \geq 3$. The luminosities are converted to absolute $V$-band magnitudes assuming $M_{V,\odot} = 4.83$ \citep{Binney1998}. This provides a sample of 163 M31 GCs.

The UCD sample that we use are the literature UCDs assembled by \cite{Norris2014}, which remains the largest sample of UCDs with measured internal velocity dispersion. For the new objects introduced in \cite{Norris2014}, we require that their calculated stellar mass, $M_*$, is below $10^8$ $M_{\odot}$. This effectively filters out larger systems (see their figure 15). We use the listed velocity dispersions, extinction-corrected absolute $V$-band magnitudes and 2D half-light radii. \cite{Norris2014} lists literature central velocity dispersions ($\sigma_0$) wherever available, which we correct to $\sigma_\infty$ with the average ratio between central and global velocity dispersions for UCDs from \cite{Mieske2008}: $\frac{\sigma_0}{\sigma_\infty} = 1.23\pm0.07$. We treat any non central $\sigma$ as approximate global values, as they are often closer to the global values than to the central values due to the compact nature of UCDs \citep{Mieske2013}. \cite{Norris2014} listed measurements for UCDs in NGC 5128 (Cen A) from \cite{Taylor2010}, which were subsequently found to be generally unreliable \citep[see][]{Mieske2013,Voggel2018,Dumont2022}, so we omit them from the sample. This gives a sample of 46 UCDs from across a variety of environments but predominantly from the Fornax and Virgo galaxy clusters.

As not all data had listed uncertainties and with the expectation that the dispersion of properties is largely driven by intrinsic differences and not by measurement uncertainties, we perform fits without considering per-object uncertainty. This is equivalent to assuming the uncertainty is uniform and equivalent to the observed intrinsic dispersion (1 standard deviation), which is $\sim0.2$ dex on $\log\sigma_\infty^2R_{h}$ and $\sim0.5$ mag on $M_V$. The total assembled dataset is available online\footnote{\url{https://github.com/bvanheumenastro/GCVDandUCDs}}.
\begin{figure}
	\includegraphics[width=\columnwidth]{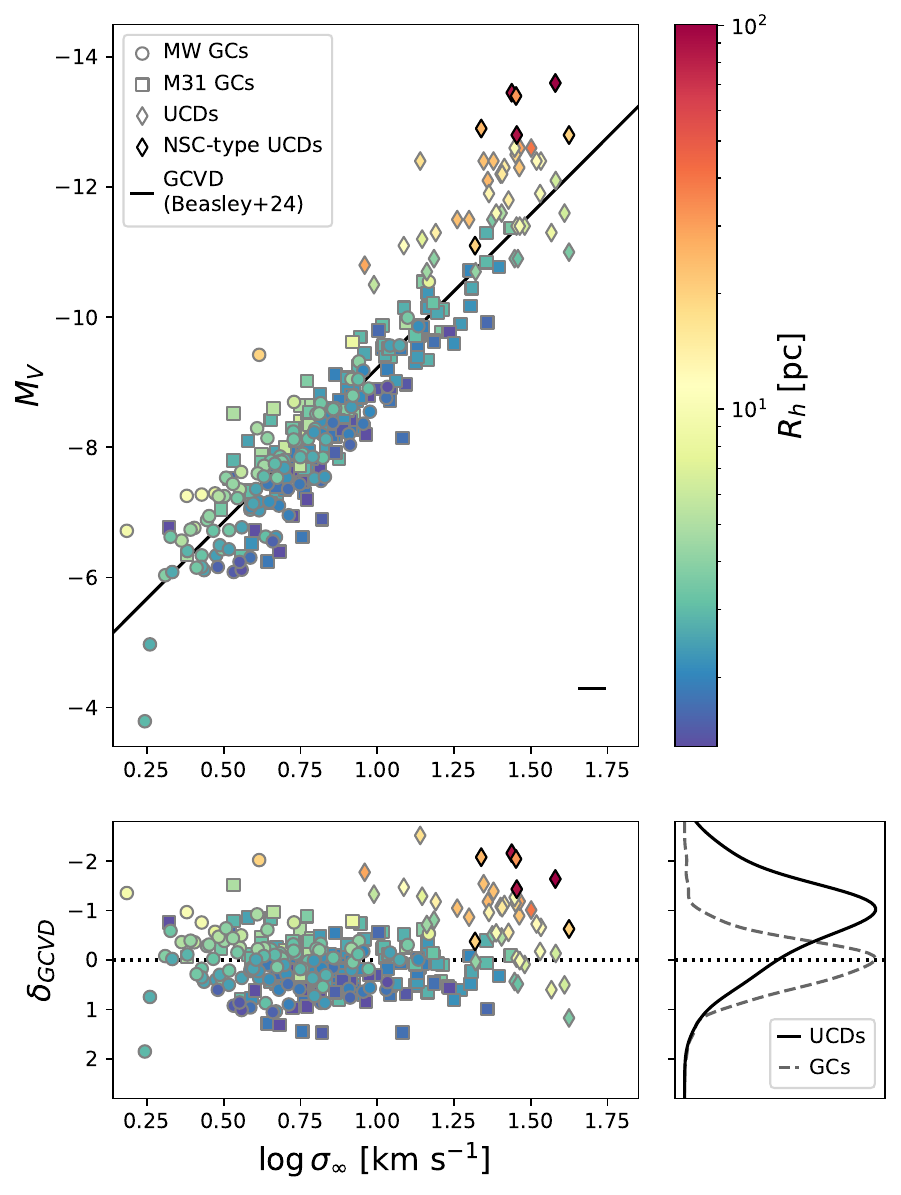}
    \caption{\textit{Top panel}: Relationship between $M_V$ and $\log\sigma_\infty$ for the MW GCs (circles), M31 GCs (squares) and UCDs (diamonds) logarithmically colour-coded by their $R_h$. The claimed stripped nuclei UCDs (see Section \ref{subsec:UCDVD}) in the sample are marked by black borders. The fiducial GCVD relation from \protect\cite{Beasley2024} is plotted as a solid black line ($M_V = -4.49 - 4.73\log\sigma_\infty$). The bar in the bottom right corner shows the difference between a central and global $\sigma$ when converted by the adopted ratio from \protect\cite{Mieske2008}. \textit{Bottom left panel}: The offsets from the GCVD ($\delta_{\rm GCVD}$) for the GCs and UCDs against $\log\sigma_\infty$ logarithmically colour-coded by $R_h$, showing a correlation between $R_h$ and $\delta_{\rm GCVD}$. \textit{Bottom right panel}: Distributions of $\delta_{\rm GCVD}$ for GCs (MW+M31; dashed gray curve) and UCDs (black curve).}
    \label{fig:GCVD_Re}
\end{figure}

\section{Analysis \& Discussion} \label{sec:results}
\subsection{The influence of $R_h$ on the GCVD} \label{subsec:GCVDoffsets}
The top panel of Figure \ref{fig:GCVD_Re} shows the $M_V-\log\sigma_\infty$ relation for the MW GCs, M31 GCs and UCDs, with the bottom panel showing the offset from the fiducial GCVD relation ($\delta_{\rm GCVD}$) of \cite{Beasley2024}. From the bottom plot of Figure \ref{fig:GCVD_Re}, it is clear that $R_h$ is the main source of the scatter on the GCVD, with the smallest GCs preferably lying below the fiducial GCVD relation while the largest GCs and UCDs lie above it. The $\delta_{\rm GCVD}$ values of the UCDs are also correlated with $\log\sigma_\infty$. This behaviour is expected as more massive UCDs, those with higher $\log\sigma_\infty$ at similar $R_h$, should show higher $M/L$ which pushes them closer to the GCVD as seen in the bottom panel of Figure \ref{fig:GCVD_Re}. Fitting the correlations for the GCs (MW + M31) and UCDs separately with linear relations gives the following relations:
\begin{equation*}
\begin{aligned}
        &\delta_{\rm GCVD, GCs} = -1.66^{+0.13}_{-0.13}\log R_h \text{[pc]} + 0.80^{+0.06}_{-0.06} \\
        &\delta_{\rm GCVD, UCDs} = -1.53^{+0.14}_{-0.15}\log R_h \text{[pc]}+ 2.63^{+0.45}_{-0.41}\log\sigma_\infty \text{[km s$^{-1}$]}\\
        &\quad\quad\quad\quad\quad\quad - 2.76^{+0.53}_{-0.57}
\end{aligned}
\end{equation*}
where the noted uncertainties are the 16 -- 84th percentile of the parameter distributions determined via bootstrapping.

The UCDs, as a population, lie systematically above the fiducial GCVD relation with an median offset of $0.72$ mag and a 1 standard deviation dispersion of $0.80$ mag. This median offset and dispersion translates to a $33.2^{+32.8}_{-36.9}$ \% underestimation for GCVD-derived distances to individual UCDs. The GCs have a dispersion of $0.53$ mag.

In the general use case of the GCVD, when the sample of objects is not limited to the brightest few, contamination from UCDs is relatively limited as UCDs are rather rare, composing only $\sim 1$\% of a typical GC system \citep[][]{Mieske2012} and the most extreme offsets are shown by the largest UCDs which are easily identified by their larger apparent $R_h$ relative to GCs. However, there is a subset of UCDs that are structurally inseparable from GCs and may show elevated $M/L$ with respect to GCs \citep[e.g.][]{Dumont2022}. Limiting the UCD sample to those with GC-like radii, $R_{h} \leq 10$ pc (16 UCDs), gives a median offset of $-0.01^{+0.18}_{-0.17}$ mag from the GCVD. The median offset is consistent with zero and below the uncertainty on the intercept of the GCVD ($\pm0.04$~mag; \citealt{Beasley2024}). This confirms the quality of the distance estimate obtained by \cite{Beasley2024} for NGC 5128 from the sample of GCs and UCDs examined by \cite{Dumont2022}. This result is due to the smooth transition in the $M/L$ -- mass relation, which means that UCDs that show GC-like structural parameters, and so similar mass, are likely to show GC-like $M/L$ and thus be more consistent with the GCVD.

The known systematics on the $R_h$ of GCs are generally minor, but may become important with the great accuracy of the GCVD. The $R_h$ of GCs increases with a shallow power-law slope of $\sim0.1-0.2$ with 2D projected galactocentric distance \citep[e.g.][]{Gomez2007,Harris2009,Webb2016} and there is a size difference of $\sim20$\% between the blue (metal-poor) and red (metal-rich) subpopulations of GCs \citep[e.g.][]{Kundu1999,Larsen2001,Spitler2006}. With a power-law slope of $0.2$, GCs located at three times the effective radius of a galaxy are $\sim24$ \% larger than those at the effective radius. For both systematics, the percentile increase translates to a difference of $\sim0.13$ mag in $\delta_{\rm GCVD}$ or $\sim6$ \% in GCVD-derived distances. This is below the determined accuracy of the GCVD and likely secondary to any stochastic variations that are intrinsic to samples.

\begin{figure*}
    \centering
    \includegraphics[width=\textwidth]{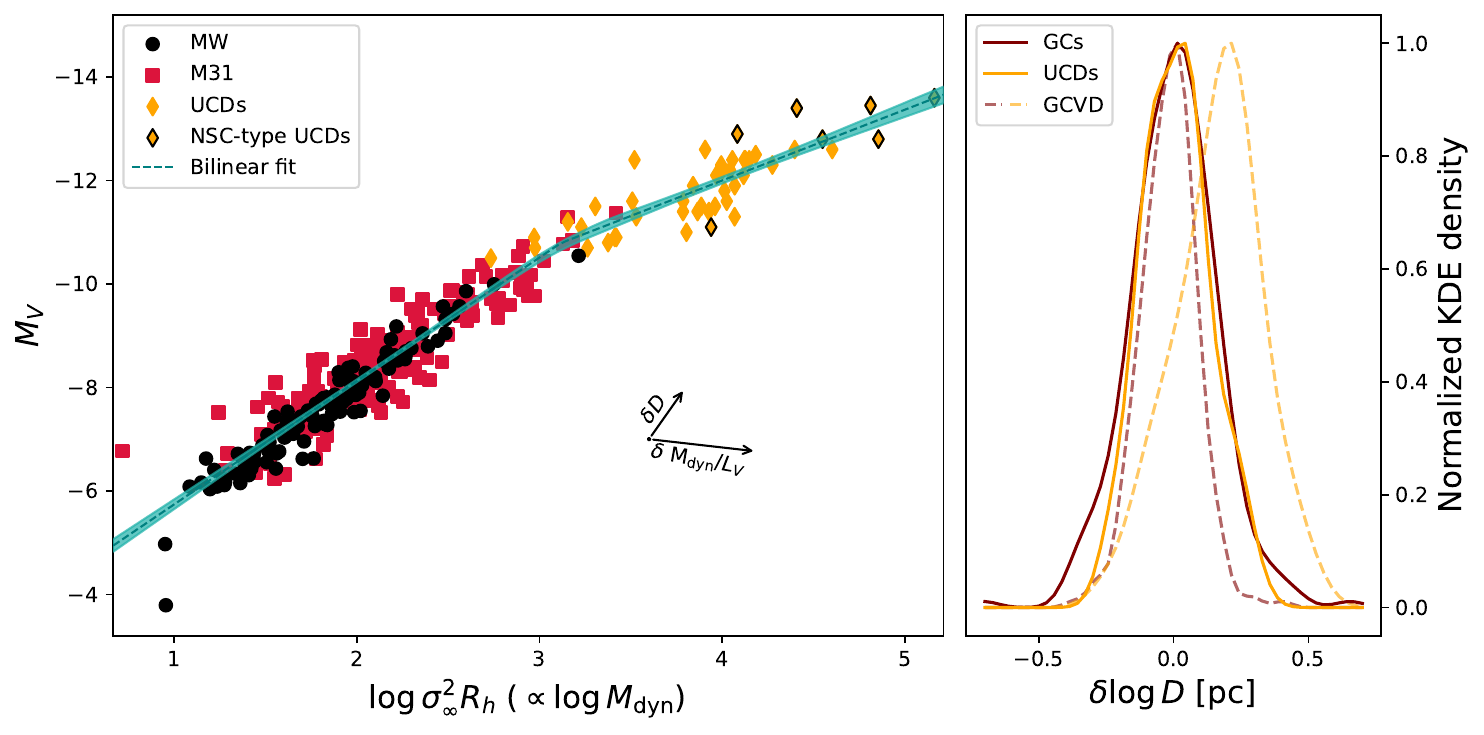}
    \caption{\textit{Left panel}: The relation between $M_V$ and $\log\sigma^2R_{h}$ for the MW GCs (black circles), M31 GCs (red squares) and UCDs (orange diamonds). Known stripped nuclei (NSC-Type) UCDs are marked with black outlines. The teal dashed line is the fitted bilinear relation with the shaded region being the 16 -- 84 percentile of the distribution of the fit parameters derived by bootstrapping. The arrows in the lower right show the directions a GC/UCD moves when it has a different $M_{\mathrm{dyn}}/L_V$ or distance. \textit{Right panel}: Kernel density estimation (KDE) distribution of the residuals on $\log D$ for the fitted bilinear relation (solid) and the GCVD (dashed) for the GCs (MW + M31; crimson) and UCDs (orange). The systematic offset for UCDs when fitting the GCVD discussed in section \ref{subsec:GCVDoffsets} is clearly visible as the dashed orange distribution is not centred at $\delta\log D = 0$.}
    \label{fig:UCDVD}
\end{figure*}

\subsection{Distances using the $M_V-\log\sigma_\infty^2R_{h}$ plane} \label{subsec:UCDVD}
The left panel of Figure \ref{fig:UCDVD} shows the relation between $\log\sigma_\infty^2R_{h}$ and $M_V$ for the MW GCs, M31 GCs and UCDs. Inspecting the panel, the relation looks as expected with two different slopes reflecting the behaviours of $M/L$ on either side of an overlapping region occurring around the expected mass breakpoint, which has been marked on the x-axis. It can be seen that the most luminous GCs, such as $\omega$Cen ($\log\sigma_\infty^2R_{h} = 3.21$), overlap strongly with the UCDs and could be considered UCDs or vice versa. We assume  a bilinear model where $x_b$ and $y_b$ are the $\log\sigma_\infty^2R_{h}$ and $M_V$ values of the breakpoint, $\alpha$ and $\beta$ are the slopes on either side of the breakpoint:
\[
M_V = 
\begin{cases}
    \alpha\log\sigma_\infty^2R_{h} + (y_b - \alpha x_b), & \text{if } \log\sigma_\infty^2R_{h} < x_b \\
    \beta\log\sigma_\infty^2R_{h} + (y_b - \beta x_b), & \text{if } \log\sigma_\infty^2R_{h} \geq x_b
\end{cases}
\]
We fit the bilinear model through linear least-squares and determine the model parameters and uncertainties through 7500 bootstrap realisations. For roughly 15\% of the bootstrap realisations, either the fitting algorithm ({\sc scipy.optimize.curve\_fit}) failed to converge, or the breakpoint was found at unreasonable values ($\log\sigma_\infty^2R_{h} \leq 1, \log\sigma_\infty^2R_{h}\geq 5$,$M_V \geq-6, M_V \leq -13$) and were discarded. This means that the bilinear model is partially degenerate with a linear model, although the distributions of the remaining realizations show the expected behaviour of a stable solution (see Figure \ref{fig:bootstrap_dists} in the Appendix). The final model parameters are quoted from the median and 16 -- 84th percentile of the remaining distributions: $x_b = 3.12^{+0.17}_{-0.16}$, $y_b = -10.79^{+0.35}_{-0.33}$, $\alpha = -2.39^{+0.08}_{-0.08}$ and $\beta = -1.37^{+0.13}_{-0.12}$. The residual dispersions for the MW GCs, M31 GCs and UCDs are 0.31 mag, 0.47 mag and 0.42 mag respectively.

The fitted slope for the GCs is just slightly shallower, but still consistent, with respect to the expectation of the Virial theorem at a constant $M/L$ ($\alpha = -2.5$). The slight tilt may reflect the slight dependence of $M/L$ on mass that is expected from their dynamical evolution \citep[e.g.][]{Mandushev1991,Kruijssen2009}. The fitted slope for the UCDs suggests $M_{\mathrm{dyn}}/L_{V} \propto M_{\mathrm{dyn}}^{0.45^{+0.05}_{-0.05}}$. The value of the breakpoint, $x_b$, translates to a mass of $\log M_{\mathrm{dyn}} [M_{\odot}] = 6.49^{+0.17}_{-0.16}$. This corresponds to a median value of $M_{\rm dyn} = 3.06\times10^6$ $M_{\odot}$, slightly larger but consistent with the mass threshold determinations from the literature. 

As both axes are distance-dependent, distance estimation with the bilinear relation is slightly more complicated than the GCVD. The bilinear relation can be rewritten to be a non-degenerate function of the logarithm of distance ($\log D$) dependent on apparent magnitude, $\log\sigma_\infty$, and apparent size ($r_h$), with objects having $m_V + 5\log\sigma_\infty^2r_h [\text{radian}] < -1.39$ lying on the `GC-branch' of the bilinear relation and those with $\geq -1.39$ on the `UCD-branch'. Based on the bootstrap realisations of the bilinear fit, the uncertainty on $\log D$ from the uncertainty of the model parameters ranges from 0.02 dex near the breakpoint to 0.06 dex at the edges of the relation. The residuals of the rearranged bilinear relation ($\delta\log D$) for both the GCs and UCDs, are shown in the right panel of Figure \ref{fig:UCDVD}. The distributions of $\delta_{\mathrm{GCVD}}$, expressed in $\delta\log D$, for GCs and UCDs are also shown in the right panel. 

The dispersion of $\delta\log D$ from the bilinear relation for GCs is 0.16 dex, larger than the 0.11 dex of the GCVD. This means that the bilinear relation requires roughly 2 times more GCs to achieve a similar statistical uncertainty. While this result seems contrary to the expectation for using dynamical mass as a tracer of luminosity, the residual dispersion between $M_V-\log\sigma_\infty^2R_h$ is smaller than that of $M_V-\log\sigma_\infty$ but the inclusion of the distance-dependent $R_h$ added another source of scatter, visually shown by the larger spread of data along the direction of shifts in distance shown by the $\delta D$ arrow in the left panel of Figure \ref{fig:UCDVD}. The dispersion for the UCDs from the bilinear relation is 0.12 dex, corresponding to a distance uncertainty of 30\% per object, and is smaller than that of the GCVD (0.16 dex) and is no longer systematically offset. With these dispersions, a minimal sample of 14 GCs or 8 UCDs is needed to reach an statistical uncertainty of 10\% on distance. Taking the total exposure time of \cite{Fahrion2026} to be a typical exposure time to obtain reliable velocity dispersion measurements for a typical GC ($\sim$4 hours) and that exposure time scales with 10$^{0.4\mu}$, where $\mu$ is the distance modulus, the per-GC integration time reaches $\sim$20 hours at a distance of $\mu =  31.5$ or 20 Mpc. As UCDs are typically brighter by $\sim$2 -- 3 mag than the typical GC, the per-UCD integration time can be a factor 6 -- 16 times lower. This means that a similar per-object integration time is achieved at $\mu = 33.5 - 34.5$ or $\sim$50 -- 80 Mpc. When applied to similar distances the observing time investment for UCDs is drastically lower, especially when multiplexing capabilities are not available.

The open question remains as to whether the composite origin of UCDs (i.e. giant/merged GCs or stripped nuclei) presents some form of systematic error. The seven claimed stripped nuclei UCDs in the sample\footnote{Fornax-UCD3 \citep[][]{Afanasiev2018}; VUCD3 \citep[][]{Ahn2017}; VUCD7 \citep[][]{Evstigneeva2008}; M59cO \citep[][]{Ahn2017}; S999 \citep[][]{Janz2015}; NGC4546-UCD1 \citep[][]{Norris2015}; Perseus-UCD13 \citep[][]{Penny2014}} (NSC-type; outlined in the left panel of Figure \ref{fig:UCDVD} with black borders) generally populate the higher $\log\sigma_\infty^2R_{h}$ values and show a similar dispersion to the UCD sample. Some of these stripped nuclei UCDs show multi-component light profiles \citep[e.g.][]{Ahn2017,Norris2015}. The \cite{Norris2014} sample adopted a model-independent value for $R_h$, resulting in these UCDs having a substantially larger $R_{h}$ than the other UCDs. We perform the same fitting procedure while excluding these UCDs and find a negligible difference in retrieved model parameters. As an additional test, we split the UCD sample into its major contributors: Fornax (20 UCDs), Virgo (13 UCDs), and treat the remaining UCDs as a group (13 UCDs) and fit a bilinear relation to each. For all three groups, we find the resulting model parameter values to be consistent with each other within their 1 standard deviation uncertainties ($\sigma_{x_b} \sim \pm0.16$, $\sigma_{y_b} \sim \pm0.35$, $\sigma_{\alpha} =\pm0.11$, $\sigma_{\beta} =\pm0.1$), although each individual fit prefers a slightly earlier breakpoint ($\langle x_b \rangle = 2.73$, $\langle y_b\rangle = -9.9$) with a steeper $\beta$, $\langle \beta\rangle = 1.63$. As the sample of UCDs that are known to be stripped nuclei or giant/merged GCs is currently very limited, we can only say that there are no hints that UCDs of different origins or environments induce systematic biases in distance estimations from the bilinear relation.

\subsection{Distance to NGC1052-DF2} \label{subsec:DF2}
\begin{figure}
    \centering
    \includegraphics[width=\columnwidth]{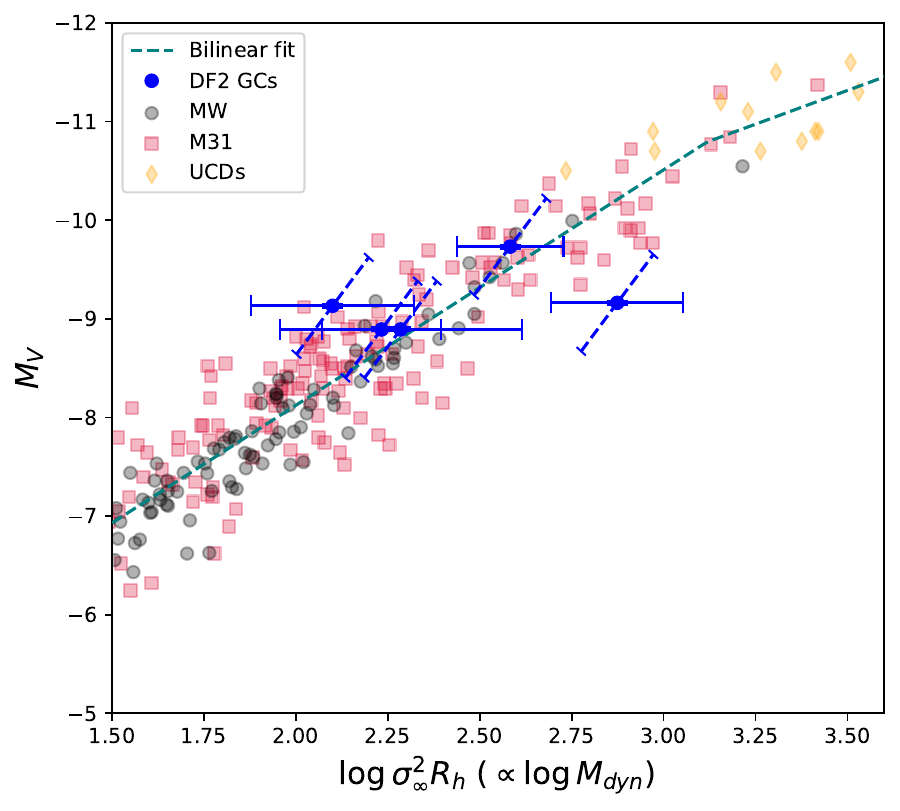}
    \caption{Position of the five brightest DF2 GCs (blue circles) in the $M_V-\log\sigma_\infty^2R_{h}$ plane. The DF2 GCs are placed at the mean derived distance of 19.1 Mpc with the dashed error bars showing the 4.3 Mpc uncertainty. The GCs and UCDs used to fit the bilinear relation are plotted with reduced opacity to show the scatter of the bilinear relation.}
    \label{fig:UCDVD_DF2}
\end{figure}

An interesting case study for this relation is the dwarf galaxy NGC1052-DF2 (hereafter DF2), whose claim of lacking dark matter \citep[][]{vanDokkum2018} and hosting large, overluminous GCs \citep[][]{vanDokkum2018gcs,Shen2021} has generated significant discourse. The key quantity in this discourse is the distance to DF2 which multiple studies, even using the same techniques, have reported different distances ranging from 13 Mpc to 21.7 Mpc \citep[e.g.][]{vanDokkum2018dist,Trujillo2019,Shen2021dist,Beasley2025,Tang2026}.

\cite{Beasley2025} measured apparent magnitude, apparent size and $\sigma_\infty$ for the five brightest clusters in DF2 for application to the GCVD. Using the same data as \cite{Beasley2025} (their table 1), we apply our fitted relation to each GC, using all bootstrap realisations of the fit and bootstrapping the measurement uncertainties assuming a Gaussian distribution ($N_{\rm bootstrap} = 500$) to define the uncertainty on individual measurements and add the dispersion of the bilinear relation in quadrature to get a total uncertainty. The distance is then calculated via a weighted mean of the individual distances. This results in a distance to DF2 of 19.1 $\pm$ 4.3 Mpc, which is consistent with the range of the literature distances within 2 standard deviations. The derived mean distance is larger than that of \cite{Beasley2025} using the same dataset, but consistent within their uncertainties. The uncertainty of the \cite{Beasley2025} estimate is, however, smaller at 3 Mpc. This is consistent with the comparison between the residuals of the bilinear relation and the GCVD for distance estimation made in Section \ref{subsec:UCDVD}.

The apparent sizes of the DF2 star clusters also lack consensus with the values from \cite{Beasley2025} being $\sim$37\% smaller than those of \cite{vanDokkum2018gcs} while the values of \cite{Trujillo2019} and \cite{Ma2020} are $\sim$10\% and $\sim$50\% larger, respectively. While the origin of the differences is not known, using the larger apparent sizes from \cite{vanDokkum2018gcs} already yields distances beyond the literature range of distances ($D_{\rm vanDokkum+18}$ = 25.1 $\pm$ 5.6 Mpc). Additionally, it has been noted that the DF2 star clusters are not generally circular \citep[e.g.][]{Trujillo2019}, which bring further uncertainty to using the circularized apparent size for mass estimation with the Virial theorem that underpins this relation.

In Figure \ref{fig:UCDVD_DF2} each of the five brightest DF2 clusters are plotted in the $M_V-\log\sigma_\infty^2R_{h}$ plane at the derived mean distance. It can be seen that all DF2 clusters lie well amongst the GCs, suggesting that they are indeed GCs and not UCDs. Furthermore GC77, the GC with the highest dynamical mass, should lie on the UCD branch of the bilinear relation but that would require a substantially larger distance (41 $\pm$ 18 Mpc), which is not supported by the other GCs. Excluding GC77 as an outlier returns a closer distance of 14.5 $\pm$ 3.8 Mpc. Within the measurement uncertainties, GC77 is however consistent with the scatter of the MW and M31 GCs. 

\section{Summary} \label{sec:summary} 
Encouraged by the excellent results achieved by the GCVD relation and the smooth transition between scaling relation behaviours of GCs and UCDs with mass, we investigated the potential of extending the GCVD to be applicable to both GCs and UCDs by including half-light radius using data for the MW GCs, M31 GCs and a catalogue of UCDs with known internal velocity dispersions. We additionally quantified the influence of UCD contamination and systematic difference in GC $R_h$ for GCVD-derived distances. 

We find that UCDs, as a population, on average lie $0.72$ mag above the fiducial GCVD with a dispersion of 0.80 mag, which corresponds to an underestimation of $\sim33^{+37}_{-33}$\% for GCVD-derived distance for UCDs. However, with a cut to select UCDs with GC-like sizes ($R_h \leq 10$~pc) the median offset is significantly reduced to $-0.01^{+0.18}_{-0.17}$ mag, fully consistent with zero. We find that the small systematic effects for GC $R_h$ will generally produce offsets smaller than the accuracy of the GCVD.

A bilinear fit to the GCs and UCDs in the $M_V-\log\sigma^2R_h$ plane results in a relation with a dispersion on $\log D$ of 0.16 dex for GCs and 0.12 dex for UCDs. This means per-object distances can be estimated to within 35\% and 30\% respectively. The breakpoint of the fitted bilinear relation suggests that the GC-UCD division occurs at $M_V \simeq -10.8$ mag and $M_{\rm dyn} \simeq 3.1\times10^{6}$ $M_{\odot}$. Splitting the UCD sample according to known/expected origin (stripped dwarf nucleus or merged GC) and their different environments, we find no hints that UCDs in either different environments or with different origins have a systematic influence on distances derived from this relation. However, the sample is currently small so that this assessment may change as more data becomes available in the future. We applied the bilinear relation to the dark matter deficient dwarf galaxy NGC 1052-DF2, which hosts overly large and overluminous GCs, and found a distance of 19.1 $\pm$ 4.3 Mpc. This is consistent with the variety of literature estimates.

\section*{Acknowledgements} 
We thank the anonymous referee for their feedback and suggestions for improving the quality of this manuscript. We further thank M. A. Beasley for the insightful discussion and the AGATE team members: D. Vaz and A. Levitskiy for their discussions and input for improving the quality of this manuscript. BvH acknowledges financial support received through a Swinburne University Postgraduate Research Award (SUPRA). DAF and JPB thank the ARC for financial support via DP220101863. AJR was supported by National Science Foundation grant AST-2308390.
\section*{Data Availability}
All data used in this work are publicly available.



\bibliographystyle{mnras}
\bibliography{example} 

\appendix
\section{Bilinear fit bootstrap distributions}
\begin{figure}
    \centering
    \includegraphics[width=\columnwidth]{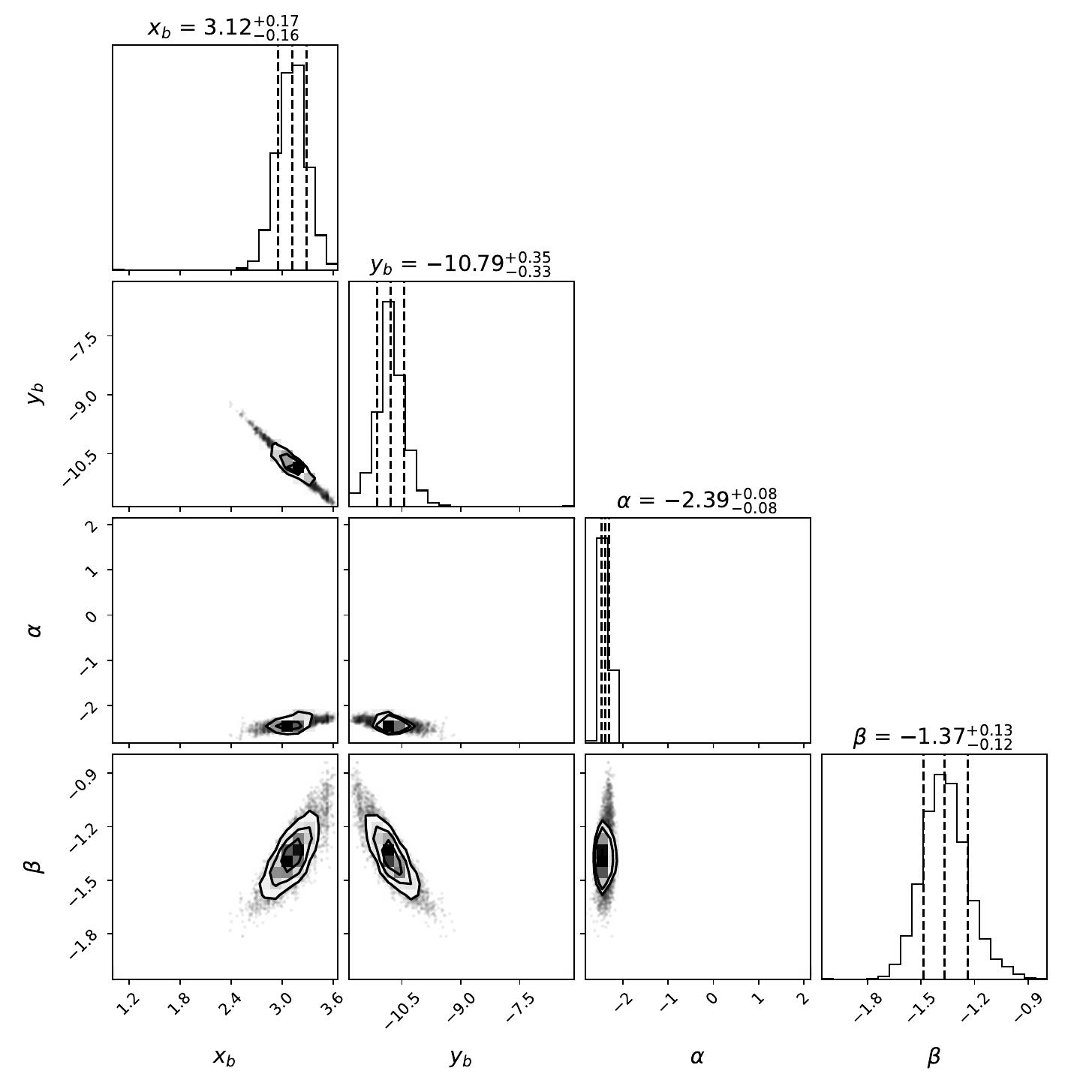}
    \caption{Corner plot of the bootstrap realisations for the fit parameters of the bilinear relation fitted to the MW GCs, M31 GCs and UCDs. The quoted numbers are the 16th, 50th and 84th percentile of the distribution of the parameters.}
    \label{fig:bootstrap_dists}
\end{figure}

Figure \ref{fig:bootstrap_dists} shows the distributions of the parameters of the bilinear relation fitted to the GCs and UCDs in the $M_V-\log\sigma^2R_h$ plane determined through bootstrapping 7500 times. Roughly 15\% of the bootstrap realizations did not converge properly and were discarded. This can be seen by the tails of the distribution in the $x_b, y_b$ plane. The parameter distributions made by the remaining bootstrap realization do show Gaussian behaviour, suggesting that the solution found is not highly degenerate.

\bsp	
\label{lastpage}
\end{document}